\documentclass[10pt,twocolumn]{IEEEtran}
\usepackage{graphicx}
\usepackage{psfrag}
\usepackage{longtable}
\usepackage{subfigure}
 
\DeclareSymbolFont{LAMSb}{U}{msb}{m}{n}
\DeclareMathSymbol{\C}{\mathalpha}{LAMSb}{"43}
\DeclareMathSymbol{\N}{\mathalpha}{LAMSb}{"4E}
\DeclareMathSymbol{\F}{\mathalpha}{LAMSb}{"46}
\DeclareMathSymbol{\R}{\mathalpha}{LAMSb}{"52}

\def\ve#1{{\mathchoice{\mbox{\boldmath$\displaystyle #1$}}%
              {\mbox{\boldmath$\textstyle #1$}}%
              {\mbox{\boldmath$\scriptstyle #1$}}%
              {\mbox{\boldmath$\scriptscriptstyle #1$}}}}

\newcommand{\BER}{{\mathrm{BER}}}
\newcommand{\FER}{{\mathrm{FER}}}
\newcommand{\MBBP}{{\mathrm{MBBP}}}
\newcommand{\transposed}{{\mathrm{T}}}

\begin{document}

\title{{\bfseries{MBBP for improved iterative channel decoding in 802.16e WiMAX systems}}}
\author{Thorsten Hehn, Johannes B. Huber, Stefan Laendner\\Chair for Information Transmission (LIT)\\University of Erlangen-Nuremberg, Germany\\$\{$hehn, huber, laendner$\}$@LNT.de}

\maketitle
\thispagestyle{empty}
\begin{abstract}We propose the application of multiple-bases belief-propagation, an optimized iterative decoding method, to a set of rate-1/2 LDPC codes from the IEEE 802.16e WiMAX standard. The presented approach allows for improved decoding performance when signaling over the AWGN channel. As all required operations for this method can be run in parallel, the decoding delay of this method and standard belief-propagation decoding are equal. The obtained results are compared to the performance of LDPC codes optimized with the progressive edge-growth algorithm and to bounds from information theory. It will be shown that the discussed method mitigates the gap to the well-known random coding bound by about 20 percent.
\end{abstract}

%\emph{Keywords: Channel Coding, Belief Propagation, LDPC codes, WiMAX IEEE 802.16e.}

\section{Introduction}
\label{sec:introduction}

The use of belief-propagation (BP) decoding \cite{pearl88} with
redundant parity-check matrix representations has drawn a lot of
attention. Several authors \cite{schwartzetal06, hanetal07,
  hollmannetal07, weberetal05, hehnetal07b} presented pioneering work
on the binary erasure channel (BEC) and provided results on the number
of redundant parity-check equations required to prevent certain
decoder failures. The concepts used on the BEC cannot be transferred
to the additive white Gaussian noise (AWGN) channel in a
straightforward manner. For this reason, several authors designed
algorithms to use redundant code descriptions for BP decoding of data
signaled over the AWGN channel. A proof of concept using the extended
Golay code of length $24$ was already given in~\cite{andrewsetal02}.
In \cite{kothiyaletal05} and \cite{jiangetal04a}, adaptive BP algorithms
were proposed. These algorithms adjust the parity-check information
for each iteration, taking into account the current decoder state.
They require additional operations which cannot be
parallelized and hence increase the delay of the data stream. The
random redundant decoding (RRD) algorithm \cite{halfordetal06} uses
multiple parity-check matrix representations in a serial fashion to
decode block codes in an iterative manner and achieves good
performance improvements. After a given number of iterations it stores
the current decoding state, changes the parity-check matrix and
resumes decoding.  For this reason it has to conduct many iterations
and thus imposes a high decoding delay. A recent paper by the authors
of the RRD algorithm \cite{halfordetal08} shows that the field of
application of this algorithm is obviously restricted to algebraic
codes.  We proposed the multiple-bases belief-propagation (MBBP)
\cite{hehnetal07} algorithm, which uses redundant parity information
in a completely parallel setting and reported good results with
algebraic codes \cite{hehnetal07}, as well as LDPC codes optimized by
the progressive edge-growth (PEG) algorithm \cite{hehnetal08, hehn09}.
In \cite{hehnetal08} we introduced the Leaking algorithm, which is a
modified BP decoding algorithm. It was shown that the combination of
MBBP and the Leaking algorithm is a valuable tool to improve the
decoding performance if a low number of redundant parity checks is
available. In this paper, we extend the field of application of MBBP
to iteratively decoded channel codes from the Worldwide
Interoperability for Microwave Access (WiMAX) standard
\cite{ieee_std_802_16e_05} and demonstrate the effectiveness of the
algorithm for this class of codes. Also, we compare the performance of
these codes to the performance of optimized PEG codes of comparable
length, both for BP and MBBP decoding.

The paper is structured as follows. In Section~\ref{sec:mbbp_decoding}
we describe the transmission setup and review the basic principles of
MBBP decoding.  Section~\ref{sec:matrix_representations} states how a
set of different parity-check matrix representations is generated, and
Section~\ref{sec:results} presents a selection of results including
the comparison to PEG codes with equal rate and similar length.
 
\section{Transmission setup and channel coding}
\label{sec:mbbp_decoding}

In this section we introduce a consistent notation and give a proper definition of the channel setup. A source emits non-redundant binary information symbols $u$, which are encoded and mapped to binary antipodal symbols $x$. As systematic encoding leads to several advantages \cite{hehnetal07e}, this type of encoding is used throughout this work. Due to the fact that exclusively $[n,k,d]$ block codes are used in this investigation, the encoded symbols are denoted as vectors $\ve{x}$ of length $n$. These vectors are transmitted over the AWGN channel. In this context, $\ve{y}$ denotes the noisy received vector corresponding to $\ve{x}$. At the receiver, an iterative decoding scheme is used to estimate $\ve{x}$ and the corresponding source symbols. This scheme is either a standard BP decoder or the MBBP decoding setup. Let us briefly review the basic properties of MBBP, which allows for the performance improvements discussed in this paper. MBBP is an iterative decoding scheme, originally designed to decode block codes with dense parity-check matrices \cite{hehnetal07c}. To this end, it runs multiple instances of the BP decoding algorithm in parallel. Each of these decoders is provided with the received signal $\ve{y}$ and a different parity-check matrix for the code. In this context, we denote the $l$ parity-check matrix representations by $\ve{H}_1$ to $\ve{H}_{l}$. The corresponding codeword estimates are $\hat{\ve{x}}_1$ to $\hat{\ve{x}}_l$ and the candidate forwarded to the information sink is $\hat{\ve{x}}$.

\begin{figure}[h!t]
\begin{center}
\psfrag{rec}[lb][lb]{$\ve{y}$}
\psfrag{success1}[l][l][1]{\scriptsize{$\ve{x}_1\ve{H}_1^{\transposed}\!=\!0$\,\,?}}
\psfrag{success2}[l][l][1]{\scriptsize{$\ve{x}_2\ve{H}_2^{\transposed}\!=\!0$\,\,?}}
\psfrag{successl}[l][l][1]{\scriptsize{$\ve{x}_l\ve{H}_l^{\transposed}\!=\!0$\,\,?}}
\psfrag{est}[c][c][1]{$\hat{\ve{x}}$}
\psfrag{BPD}[c][c][1]{{\scriptsize{BP-Dec.}}}
\psfrag{on}[c][c][1]{\scriptsize{{on}}}
\psfrag{c1}[l][l]{\scriptsize{$\hat{\ve{x}}_1$}}
\psfrag{c2}[l][l]{\scriptsize{$\hat{\ve{x}}_2$}}
\psfrag{cl}[l][l]{\scriptsize{$\hat{\ve{x}}_{l_{\MBBP}}$}}
\psfrag{Rep1}[c][c][1]{{\scriptsize{$\ve{H}_1$}}}
\psfrag{Rep2}[c][c][1]{\scriptsize{{$\ve{H}_2$}}}
\psfrag{Repl}[c][c][1]{{\scriptsize{$\ve{H}_{l_{\MBBP}}$}}}
\psfrag{dots}[c][c][1]{{\scriptsize{$\dots$}}}
\psfrag{PostProc}[c][c][1]{{\scriptsize{Post-Proc.}}}
\psfrag{I}[lb][lb][1]{{\scriptsize{$I_{\max,\MBBP}$}}}
\includegraphics[scale=0.4]{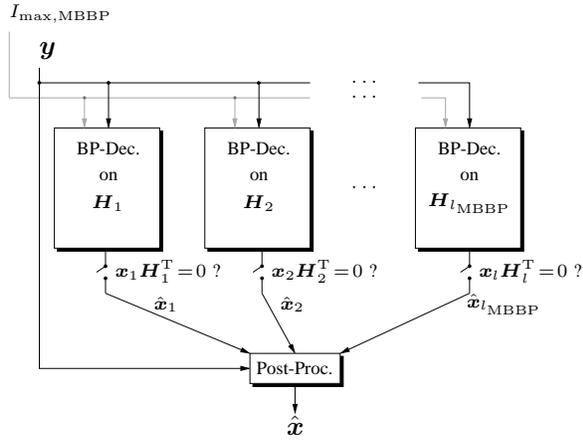}
\caption{MBBP decoding setup}
\label{fig:MBBP_NX_S}
\end{center}
\vspace{-0.7cm}
\end{figure}

This algorithm is motivated by the fact that different parity-check matrices allow for decoding of different error patterns when the suboptimal BP decoding algorithm is used. This can be understood as ``decoder diversity''. It is the task of the post-processing unit to combine these capabilities such that one single decoder estimate can be forwarded to the information sink. This is a degree of freedom in the MBBP design. Moreover, the opportunity of allowing the decoders to communicate with each other exists.
A multitude of variations of this type have been introduced in~\cite{hehnetal07c}. In this work, we focus on non-communicating parallel BP decoders and a post-processing unit that deploys the usual Euclidean distance metric. It uses this metric to select the codeword which is closest to the received vector $\ve{y}$ from the decoder outputs (MBBP-NX-S setup in \cite{hehnetal07c}). Figure~\ref{fig:MBBP_NX_S} visualizes this approach.

We will also make use of the Leaking algorithm to improve the decoding performance of WiMAX LDPC codes. This method keeps channel information from the decoder and allows it only to ``leak'' into the decoding progress with rising iteration number. It was shown that this method mitigates the problems of BP decoding with short cycles. In order to use this approach, two variables need to set: $p_{\mathcal{L}}$, the probability of a variable node being informed on the channel output in the first iteration as well as the parameter $I'_{\mathrm{max}}$, the (hypothetical) iteration number for which all channel information is included in the decoding process. Detailed information on this algorithm can be found in \cite{hehnetal08}. We refer to an MBBP setup using the Leaking algorithm by L-MBBP.

The WiMAX standard provides a multitude of channel codes. A performance comparison for signaling over the AWGN channel, confirming that the class of LDPC codes is among the most powerful codes in this setup, can be found in \cite{baumgartneretal05}. Inspired by its short length and a discussion on the practical relevance of these codes \cite{kienle08}, the rate-$1/2$ LDPC codes proposed in the IEEE 802.16e standard \cite{ieee_std_802_16e_05} are investigated in this work. As we are generally interested in codes of short length, we restrict our attention to codes of length $576\leq n\leq 960$. For comparison we also consider PEG-optimized codes of rate $1/2$ and length $500\leq n\leq 1000$.

\section{Parity-check matrix representations}
\label{sec:matrix_representations}

We review a general method to construct redundant parity-checks from a given matrix and present a novel method for LDPC codes from the WiMAX standard. A comparison shows the advantage of the second approach.

\subsection{Set of matrix representations for MBBP decoding}
\label{sec:matrix_representations_for_mbbp}

The most crucial parameter for the success of MBBP decoding is the set of parity-check matrices used in the decoding instances. Especially simulations with PEG codes \cite{hehnetal08} have shown that the applied matrices need to fulfill two criteria. First, the Tanner graphs \cite{tanner81} of the matrices need to differ sufficiently in their structure such that the decoders obtain a decoding diversity and a performance improvement. Second, the decoders running on the parity-check matrices need to obtain comparable performance results. Adding a representation to an existing MBBP setup can only increase the overall performance if its standard BP performance is comparable to the performance of the current MBBP setup.

In \cite{hehnetal08} a general method to construct a set of redundant parity-check matrices for a given code was presented. This method was originally intended to provide good redundant parity-check matrices for PEG-constructed codes of short length and makes use of the fact that there exist cycles of length $4$ and $6$ in the Tanner graph. Especially for code lengths $n\leq 1000$, many additional parity-check equations can be found with this method. The aim of this approach, to be described in detail shortly, is to approximate the property ``low-density'' for the additional checks. Let $c$ be the length of the considered cycle and let ${\mathcal{G}}_c$ be one set of indices of parity checks closing a cycle of length $c$. A linear combination of the parity checks indexed by the set ${\mathcal{G}}_c$ leads to a novel parity-check equation with a (Hamming) weight of at most

\begin{equation}
w_{\mathrm{r}}=\sum\limits_{i\in{\mathcal{G}}_c} w_i - c,
\label{eq:weight_redundant_row}
\end{equation}
where $w_i$ denotes the weight of parity check $i$. This is a general approach and can be used for any parity-check matrix with a local cycle length of $c$. It was shown in \cite{hehnetal08} and \cite{hehn09} that this approach leads to desirable performance results when using PEG codes. However, the parity-check matrix of the WiMAX codes show more structure, what allows for a better construction algorithm.

\subsection{Parity-check matrices for codes specified in the IEEE 802.16e standard}
\label{sec:matrix_representations_80216e}

The LDPC codes of rate $1/2$ standardized in \cite{ieee_std_802_16e_05} are all deducted from one base matrix $\ve{H}'_{\mathrm{b}}$. The realizations of different lengths are created from this matrix by {\emph{lifting}} \cite{tanner81}. Prior to this step, a renormalization is done, i.e.\ the lifting procedure is applied to the elements

\begin{equation}
H_{\mathrm{b}}(i,j)=\left\{\begin{array}{ccc}\left\lfloor\frac{H'_{\mathrm{b}}(i,j)\cdot z}{96}\right\rfloor&\mbox{ if }& H'_{\mathrm{b}}(i,j)>0\\H'_{\mathrm{b}}(i,j)&\mbox{ if }&H'_{\mathrm{b}}(i,j)\leq 0\end{array}\right.
\end{equation}
of the matrix $\ve{H}_{\mathrm{b}}$.

\begin{figure*}
\begin{equation}
\arraycolsep0.65mm
\ve{H}'_{\mathrm{b}}=\left( \begin{array}{rrrrr rrrrr rrrrr rrrrr rrrr}
-1 & 94 & 73 & -1 & -1 & -1 & -1 & -1 & 55 & 83 & -1 & -1 & 7 & 0 & -1 & -1 & -1 & -1 & -1 & -1 & -1 & -1 & -1 & -1  \\ 
-1 & 27 & -1 & -1 & -1 & 22 & 79 & 9 & -1 & -1 & -1 & 12 & -1 & 0 & 0 & -1 & -1 & -1 & -1 & -1 & -1 & -1 & -1 & -1  \\ 
-1 & -1 & -1 & 24 & 22 & 81 & -1 & 33 & -1 & -1 & -1 & 0 & -1 & -1 & 0 & 0 & -1 & -1 & -1 & -1 & -1 & -1 & -1 & -1  \\ 
61 & -1 & 47 & -1 & -1 & -1 & -1 & -1 & 65 & 25 & -1 & -1 & -1 & -1 & -1 & 0 & 0 & -1 & -1 & -1 & -1 & -1 & -1 & -1  \\ 
-1 & -1 & 39 & -1 & -1 & -1 & 84 & -1 & -1 & 41 & 72 & -1 & -1 & -1 & -1 & -1 & 0 & 0 & -1 & -1 & -1 & -1 & -1 & -1  \\ 
-1 & -1 & -1 & -1 & 46 & 40 & -1 & 82 & -1 & -1 & -1 & 79 & 0 & -1 & -1 & -1 & -1 & 0 & 0 & -1 & -1 & -1 & -1 & -1  \\ 
-1 & -1 & 95 & 53 & -1 & -1 & -1 & -1 & -1 & 14 & 18 & -1 & -1 & -1 & -1 & -1 & -1 & -1 & 0 & 0 & -1 & -1 & -1 & -1  \\ 
-1 & 11 & 73 & -1 & -1 & -1 & 2 & -1 & -1 & 47 & -1 & -1 & -1 & -1 & -1 & -1 & -1 & -1 & -1 & 0 & 0 & -1 & -1 & -1  \\ 
12 & -1 & -1 & -1 & 83 & 24 & -1 & 43 & -1 & -1 & -1 & 51 & -1 & -1 & -1 & -1 & -1 & -1 & -1 & -1 & 0 & 0 & -1 & -1  \\ 
-1 & -1 & -1 & -1 & -1 & 94 & -1 & 59 & -1 & -1 & 70 & 72 & -1 & -1 & -1 & -1 & -1 & -1 & -1 & -1 & -1 & 0 & 0 & -1  \\ 
-1 & -1 & 7 & 65 & -1 & -1 & -1 & -1 & 39 & 49 & -1 & -1 & -1 & -1 & -1 & -1 & -1 & -1 & -1 & -1 & -1 & -1 & 0 & 0  \\ 
43 & -1 & -1 & -1 & -1 & 66 & -1 & 41 & -1 & -1 & -1 & 26 & 7 & -1 & -1 & -1 & -1 & -1 & -1 & -1 & -1 & -1 & -1 & 0  \\ 
\end{array}\right)
\label{eq:base_matrix}
\end{equation}
\end{figure*}

In this context, $z$ is the \emph{expansion factor} and depends on the code realization.
The lifting procedure, from which the parity-check matrix $\ve{H}$ results, is described as follows. Each negative entry in the base matrix $\ve{H}_{\mathrm{b}}$ is replaced by a $z\times z$ zero matrix and each non-negative element $H_{\mathrm{b}}(i,j)$ is substituted by an identity matrix which is cyclically shifted to the right by $H_{\mathrm{b}}(i,j)$ positions.  Equation~(\ref{eq:base_matrix}) specifies the base matrix $\ve{H}_{\mathrm{b}}'$ for the rate-$1/2$ LDPC code \cite[p.\ 628]{ieee_std_802_16e_05}.

Performing the lifting approach leads to the binary matrix $\ve{H}$. 
Considering that any entry in $\ve{H}_{\mathrm{b}}$ is replaced by a permutation matrix with constant row weight one, it is easy to see from Equation~(\ref{eq:base_matrix}) that the weight of any parity check of $\ve{H}$ is $6$ or $7$, regardless of the actual length $n$ of the code. The girth of the code was found to be $6$ for all lengths considered. It is now our task to determine parity checks which are linear combinations of the given parity checks and have as low as possible weight.

\subsection{Redundant parity checks from $\ve{H}_{\mathrm{b}}$}
\label{sec:redundant_checks_from_base_matrix}

The novel approach for creating redundant parity-check equations uses the base matrix $\ve{H}_{{\mathrm{b}}}$ instead of the binary matrix $\ve{H}$ to find valid linear combinations. Subsequently it performs the lifting operation on the redundant checks. Let us elaborate on the generation of these checks. In a binary matrix, a redundant check can be found as a linear combination of two or more existing checks. This proceeding is in general not possible when the base matrix $\ve{H}_{\mathrm{b}}$ is considered, as the addition of two entries is not defined. However, the addition of a negative and a non-negative element, as well as the addition of two zero elements is a straightforward task. The result of the addition is the non-negative element and the element $-1$, respectively. Using this approach, redundant checks can be created by the linear combination of two existing checks, which do not share a positive element in any column. Lifting a redundant check leads to a set of $z$ checks for the binary matrix $\ve{H}$, which are subsequently used to create sets of non-equal, binary parity check matrices. As an example, we state that the linear combination of rows $11$ and $12$ in $\ve{H}_{\mathrm{b}}$ leads to $z$ binary redundant checks of weight $10$, since the non-negative entries in rows $11$ and $12$ have disjoint column positions, except for the last column, which contains zero entries.

Depending on the length of the code, we replace $10$ to $16$ parity checks in the existing parity-check matrix to generate a new representation. At this step, we ensure that the resulting parity-check matrix has full rank.

%\footnote{If the additional matrix representations are created in such a way that all $z$ binary checks emanating from one redundant check of $\ve{H}_{\mathrm{b}}$ are deployed, the resulting binary matrix preserves the structural properties of the original parity-check matrix, i.e.\ it consists only of $z\times z$ zero matrices and cyclically shifted identity matrices of the same size.} 

Let us now compare this result to the approach from Section~\ref{sec:matrix_representations_for_mbbp} by means of the WiMAX code of length $n=576$. The local girth of its parity-check matrix varies between $c=6$ and $c=8$. Using Equation~(\ref{eq:weight_redundant_row}) and $c=6$, it can be deducted that additional parity-check representations have a weight of at most $12$ to $15$, depending on the actual parity checks used to create the linear combinations.
It was verified by computer simulations that this bound is met by the realizations.

The authors are aware of the fact that this novel method is still a suboptimal approach and therefore assess it in a more general manner.
The methods provided in \cite{Huetal04} allow for an efficient search of low-weight codewords. Using these methods to the dual of the IEEE 802.16e rate-$1/2$-code of length $576$ did not return any codewords which are not already present in the rows of the original parity-check matrix of weight below $10$. This allows us to conclude that the proposed method is well suited for the considered class of codes.

\section{Results and Comparison}
\label{sec:results}

We present simulation results for codes from the WiMAX standard of different length. In this context, we apply standard BP decoding as well as the (L)-MBBP approach to show the performance improvements obtained with this method. We allow all BP decoding units to perform at most $200$ iterations. This is a proper choice for which a further increase does not improve the standard BP decoding performance significantly. We also limit the number of different parity-check matrices in an MBBP setup to 15 and allow leaking with an initial setting of $p_{\mathcal{L}}=0.9$, what results to a maximum number of $30$ decoders in parallel. The current development of multiprocessor techniques \cite{vangaletal07} allows us to state that this setting can easily be parallelized with upcoming microcontroller techniques.
Furthermore we set the parameter $I'_{\mathrm{max}}=300$, as this setting leads to desirable results in our computer simulations.

\begin{figure}[h!t]
\begin{center}
\subfigure{
\psfrag{SNR}[ct][ct]{{\small{$10\log_{10}(E_b/N_0)\,\rightarrow$}}}
\psfrag{BER}[cb][cb]{{\small{$\BER\,\rightarrow$}}}
\psfrag{BP, 200}[l][l]{{\scriptsize{BP}}}
\psfrag{MBBP set 1}[l][l]{{\scriptsize{MBBP, $l=7$}}}
\psfrag{MBBP set 2}[l][l]{{\scriptsize{MBBP, $l=15$}}}
\psfrag{Leaking-MBBP}[l][l]{{\scriptsize{L-MBBP, $l=30$}}}
\psfrag{Gallager bound}[l][l]{{\scriptsize{Gallager bound}}}
\psfrag{n=576}[l][l]{{\scriptsize{$n=576$}}}
\psfrag{n=960}[l][l]{{\scriptsize{$n=960$}}}
\includegraphics[scale=0.63]{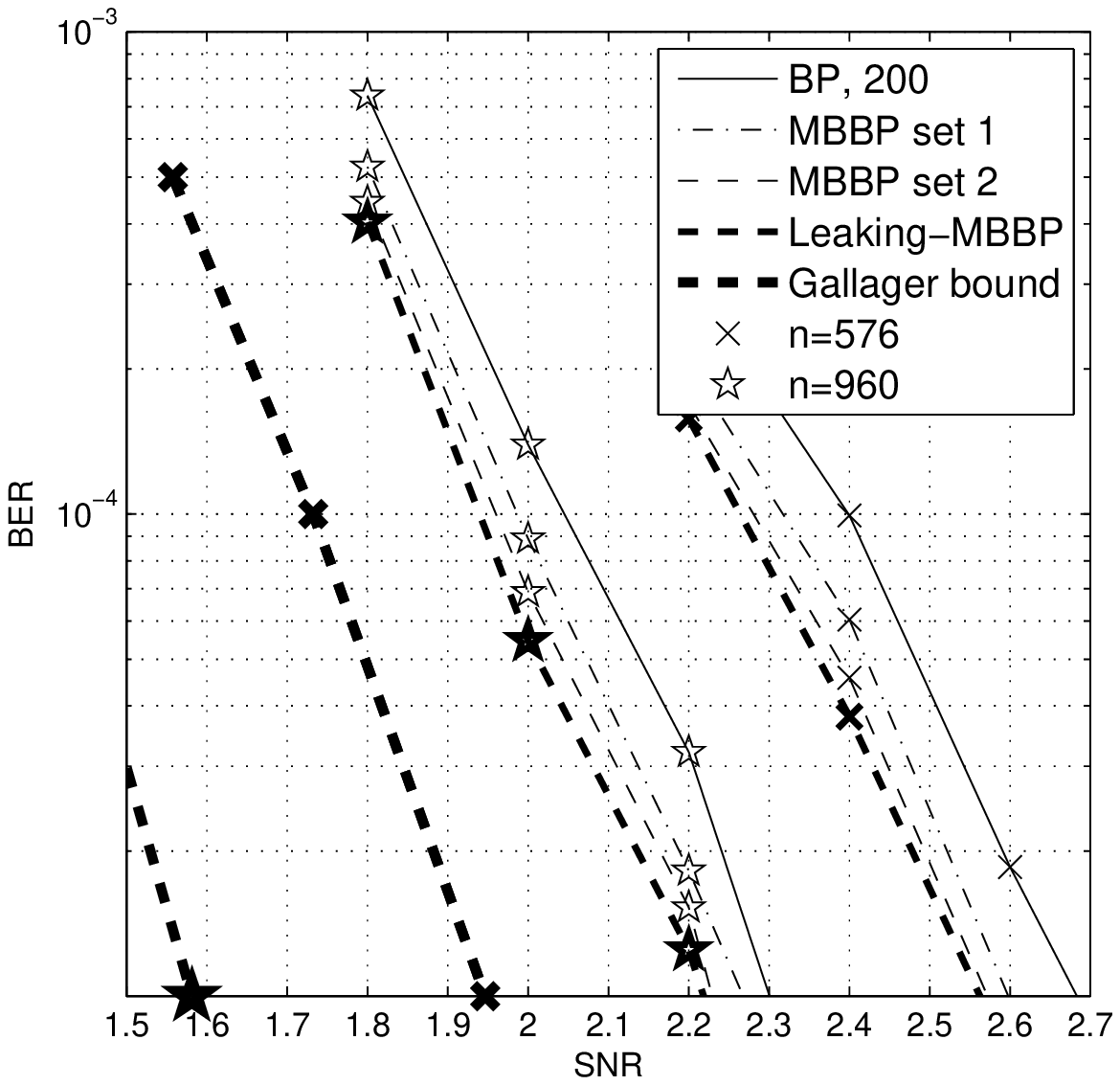}
}
\subfigure{
\psfrag{SNR}[ct][ct]{{\small{$10\log_{10}(E_b/N_0)\,\rightarrow$}}}
\psfrag{FER}[cb][cb]{{\small{$\FER\,\rightarrow$}}}
\psfrag{BP, 200}[l][l]{{\scriptsize{BP}}}
\psfrag{MBBP set 1}[l][l]{{\scriptsize{MBBP, $l=7$}}}
\psfrag{MBBP set 2}[l][l]{{\scriptsize{MBBP, $l=15$}}}
\psfrag{Leaking-MBBP}[l][l]{{\scriptsize{L-MBBP, $l=30$}}}
\psfrag{Gallager bound}[l][l]{{\scriptsize{Gallager bound}}}
\psfrag{n=576}[l][l]{{\scriptsize{$n=576$}}}
\psfrag{n=960}[l][l]{{\scriptsize{$n=960$}}}
\includegraphics[scale=0.63]{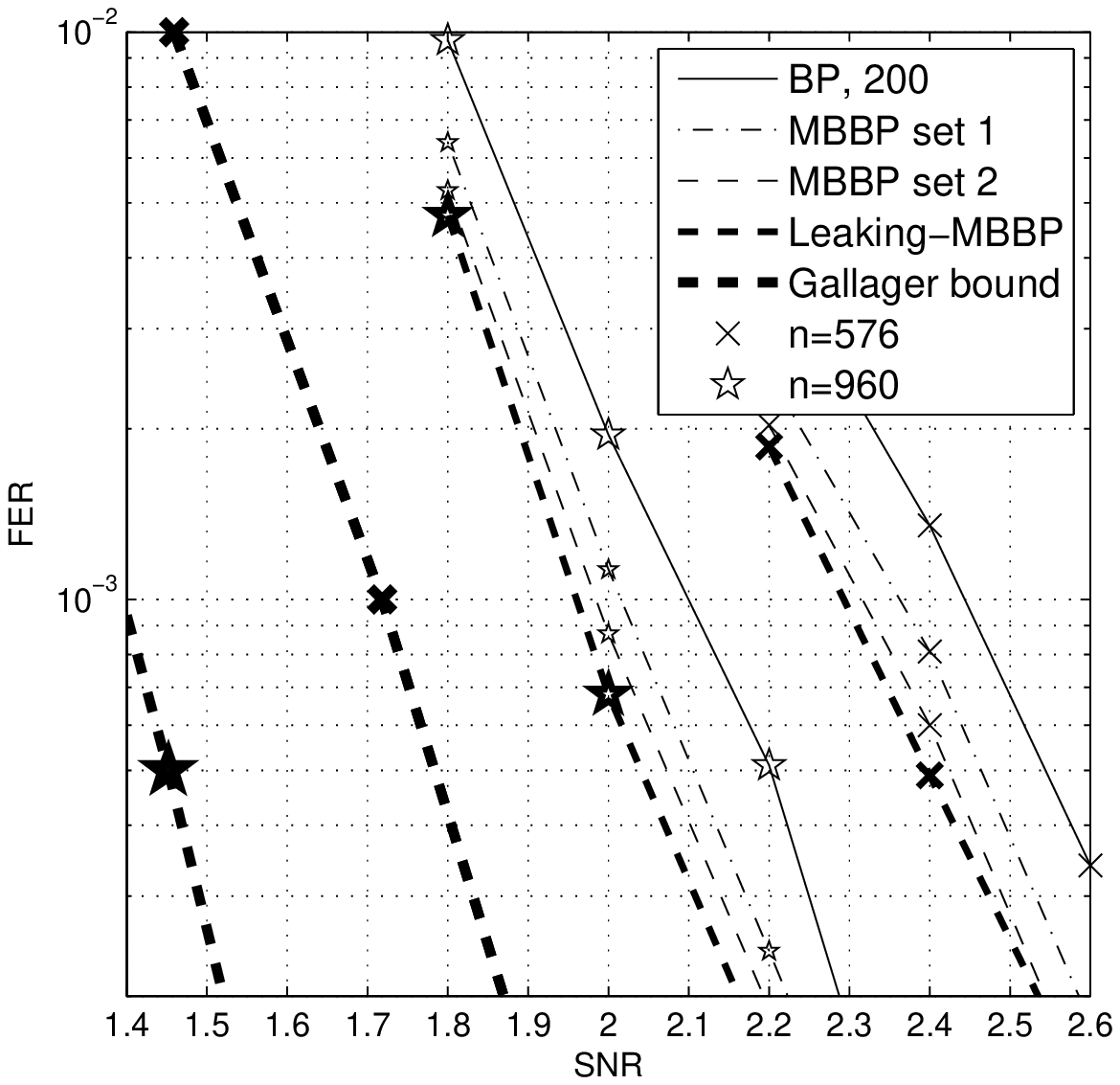}
}
\caption{\label{fig:ieee_80216_15_decoders}BER and FER performance comparison of IEEE 802.16e codes with BP and the L-MBBP approach.}
\end{center}
\vspace{-0.3cm}
\end{figure}

Figure~\ref{fig:ieee_80216_15_decoders} shows performance results for the WiMAX codes of length $n=576$ and $n=960$. In order to emphasize that the bigger part of the decoding gain is already obtained by a low number of decoder representations, we show different MBBP settings. To be precise, we allow $l=7$, $l=15$, and $l=30$ representations to run in parallel. In Figure~\ref{fig:ieee_80216_15_decoders} we observe that the most prominent part of the decoding gain is already achieved with $7$ decoders in parallel and another small gain is achieved for $l=15$. The setup using L-MBBP and utilizing $30$ decoders in total compares favorably but the difference is small in relation to the number of decoders additionally required. Using all decoding units, the proposed multi-decoding approach improves the performance of WiMAX codes for about $0.15$ dB. The random coding bound (Gallager bound) \cite{gallager68} marks desirable performance values and is shown for comparison reasons. In order to provide performance results on the $\BER$, we estimate the minimum distance $d$ for random codes of given length and rate by means of the Gilbert-Varshamov-bound \cite{macwilliamsetal77} and assume for the random coding bound that $d$ errors happen in an erroneously decoded frame. Details on this approach can be found in~\cite{hehnetal07e}. 

In a next step, we assess our results in a more general setting and compare the codes defined in the WiMAX standard to PEG-optimized codes of equal rate and length $500\leq n \leq 1000$. We also compare these results to the Gallager bound and the sphere packing bound. Detailed information on the latter bound can be found in \cite{shannon59}. For the PEG codes, we use the optimized degree distribution

\begin{eqnarray}
L(x)&=&0.5043865558\cdot x^2+0.2955760529\cdot x^3+\nonumber\\&&0.0572634080\cdot x^5+0.0362602194\cdot x^6+\nonumber\\&&0.0049622081\cdot x^7+0.0292344776\cdot x^9+\nonumber\\&&0.0650312477\cdot x^{11}+0.0072858305\cdot x^{12}
%\\
%R(x)&=&1\cdot x^7
\label{eq:degree_distribution}
\end{eqnarray}
from \cite{urbanke}, which has a gap to capacity of about $0.2$ dB and leads to desirable results for the code lengths of interest \cite{hehnetal07e}. Within the error region of interest, the created ensembles show strictly concentrated behavior, what allows us to study the subsequent results independent of the random seed used for the construction algorithm.

\begin{figure*}[htb]
\begin{minipage}{7.5cm}
\psfrag{Length}[ct][ct]{{\small{Length $n\,\rightarrow$}}}
\psfrag{Required EbN0 in dB}[cb][cb]{{\small{Required $10\cdot\log_{10}(E_{\mathrm{b}}/N_0)\,\rightarrow$}}}
\psfrag{BP}[l][l]{{\scriptsize{BP}}}
\psfrag{L-MBBP}[l][l]{{\scriptsize{L-MBBP}}}
\psfrag{Gallager}[l][l]{{\scriptsize{Gallager}}}
\psfrag{SPB}[l][l]{{\scriptsize{SPB}}}
\psfrag{BER con}[c][c]{\small{$\BER=10^{-5}$}}
\includegraphics[scale=0.6]{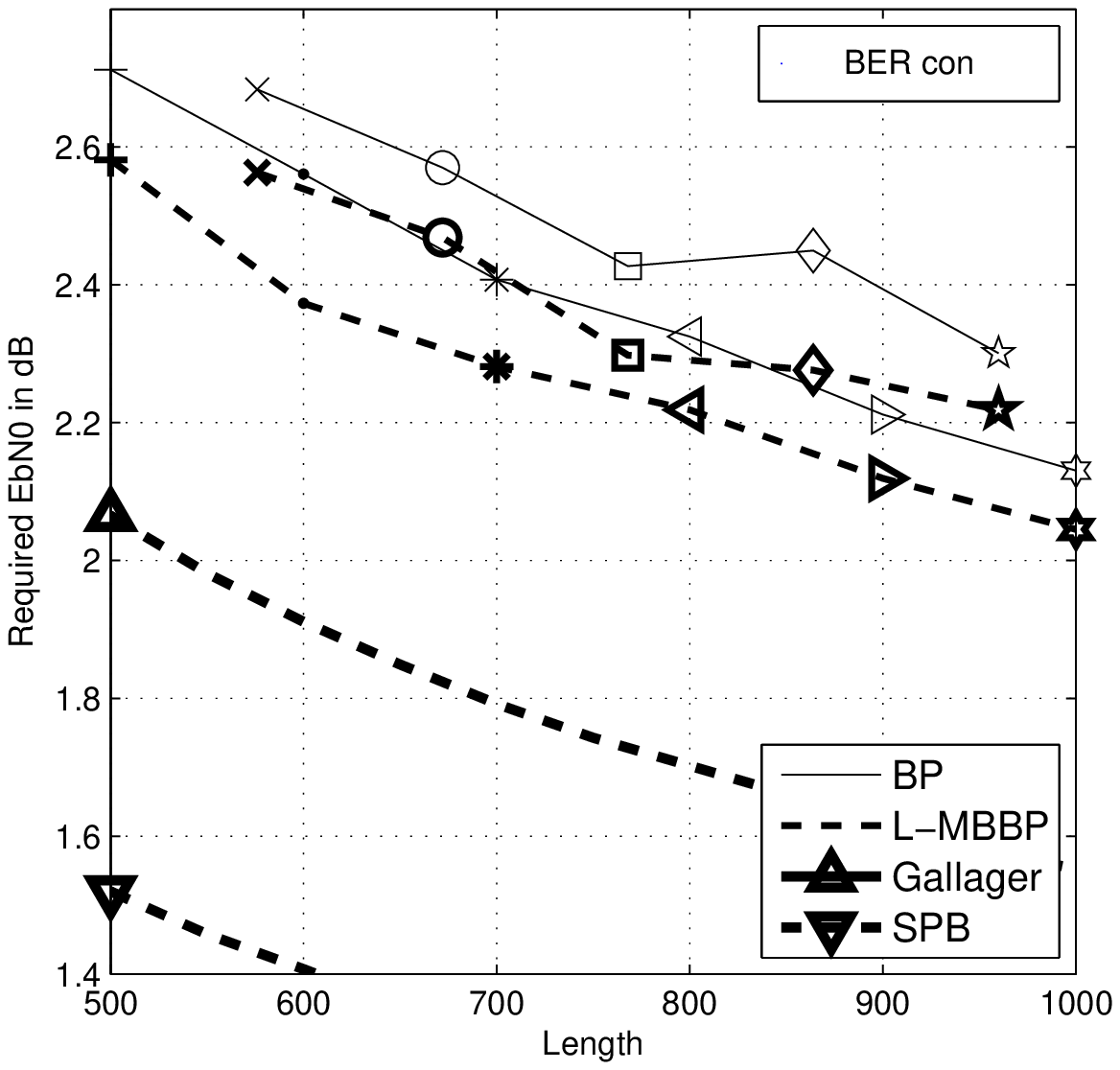}
\end{minipage}\begin{minipage}{7.5cm}
\psfrag{Length}[ct][ct]{{\small{Length $n\,\rightarrow$}}}
\psfrag{Required EbN0 in dB}[cb][cb]{{\small{Required $10\cdot\log_{10}(E_{\mathrm{b}}/N_0)\,\rightarrow$}}}
\psfrag{BP}[l][l]{{\scriptsize{BP}}}
\psfrag{L-MBBP}[l][l]{{\scriptsize{L-MBBP}}}
\psfrag{Gallager}[l][l]{{\scriptsize{Gallager}}}
\psfrag{SPB}[l][l]{{\scriptsize{SPB}}}
\psfrag{FER con}[c][c]{\small{$\FER=10^{-3}$}}
\includegraphics[scale=0.6]{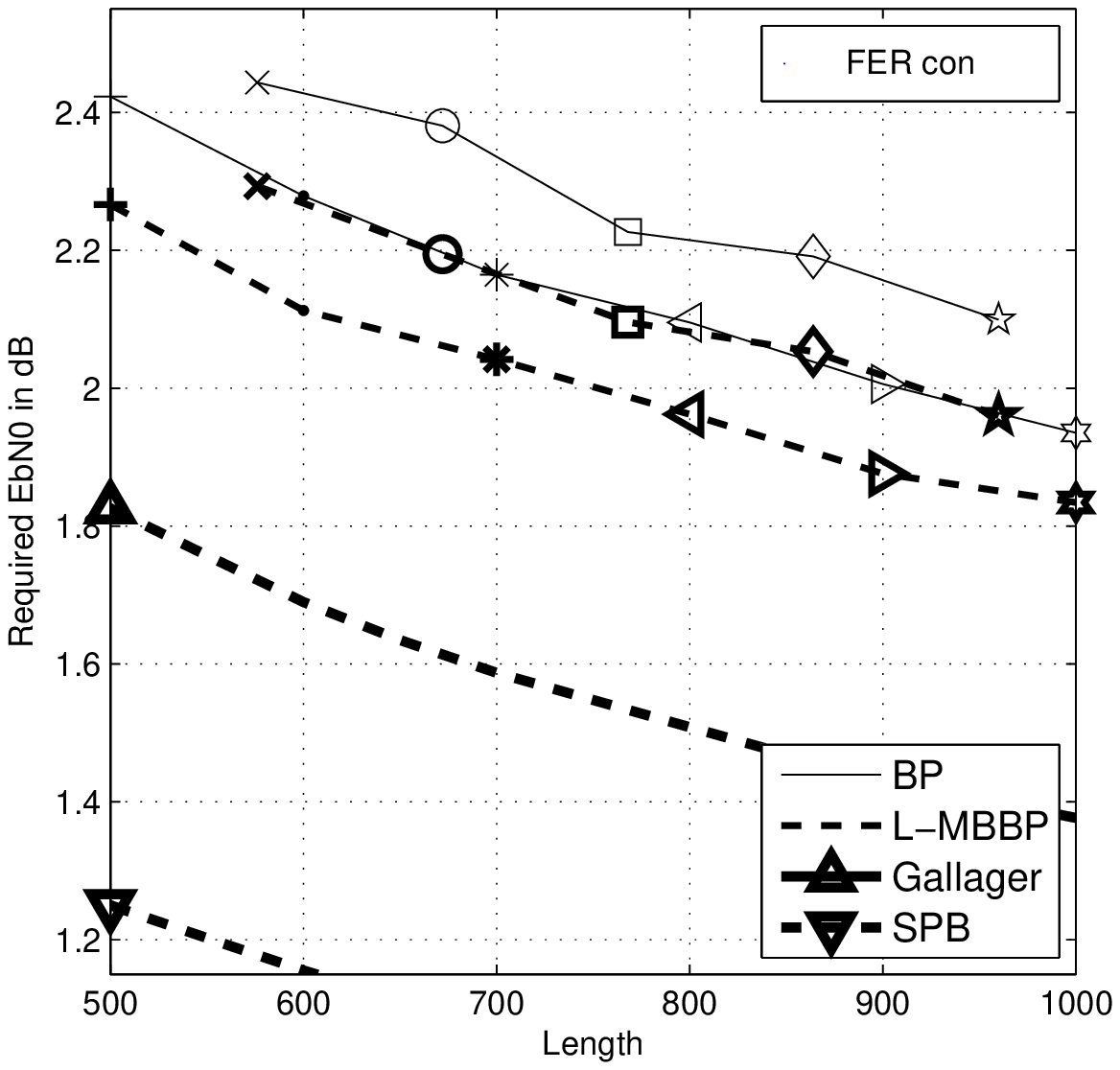}
\end{minipage}\begin{minipage}{3cm}
\psfrag{Wimax}[l][l]{\underline{WiMAX}}
\psfrag{n=576}[l][l]{$n=576$}
\psfrag{n=672}[l][l]{$n=672$}
\psfrag{n=768}[l][l]{$n=768$}
\psfrag{n=864}[l][l]{$n=864$}
\psfrag{n=960}[l][l]{$n=960$}
\psfrag{PEG}[l][l]{\underline{PEG}}
\psfrag{n=500}[l][l]{$n=500$}
\psfrag{n=600}[l][l]{$n=600$}
\psfrag{n=700}[l][l]{$n=700$}
\psfrag{n=800}[l][l]{$n=800$}
\psfrag{n=900}[l][l]{$n=900$}
\psfrag{n=1000}[l][l]{$n=1000$}
\includegraphics[scale=0.73]{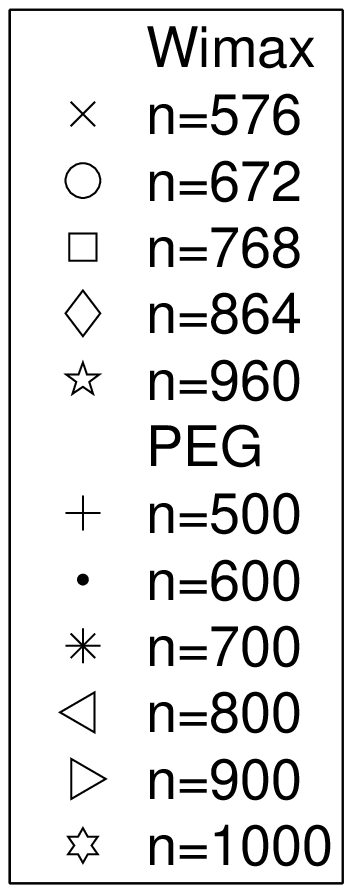}
\end{minipage}
\caption{\label{fig:wimax_and_peg_required_SNR_over_codelength}Required SNR to meet given quality constraints $\BER=10^{-5}$ and $\FER=10^{-3}$, respectively, for WiMAX codes ($576\leq n \leq 960$) and PEG-optimized codes ($500\leq n \leq 1000$). For comparison, the Gallager bound and the SPB are shown.}
\end{figure*}

Figure~\ref{fig:wimax_and_peg_required_SNR_over_codelength} shows the signal-to-noise ratio {$10\cdot\log_{10}(E_{\mathrm{b}}/N_0)$}, which is required to obtain the reliability criterion $\BER=10^{-5}$ and $\FER=10^{-3}$, respectively. Plotted are results for WiMAX codes and PEG-optimized codes for both BP and L-MBBP decoding as well as the Gallager bound. In order to keep an appropriate presentation for the numerical results, but still provide the reader with an idea of the position of the sphere packing bound (SPB), we choose to plot the left-most part of it and state that its shape is shown to be similar to the shape of the Gallager bound in~\cite{hehnetal07e}.

Let us first consider the results for the WiMAX codes. In our simulations, the codes showed slightly different error-floor behavior. This holds in particular for the code of length $n=864$. It is to observe that a gain of about $0.15$ dB is achieved for all code lengths considered. From the plot for $\FER=10^{-3}$ and the code of length $n=960$ we observe that the gap to the Gallager bound reads about $0.7$ dB, which can be lowered by $0.14$ dB (or $20$ $\%$) with the L-MBBP approach.

Similar results are presented for the PEG-optimized LDPC codes, where we also restrict the maximum number of decoders in parallel to $30$. The actual number is however lower due to lack of well-performing presentations.
The PEG codes show the desired performance results at about $0.15$ dB lower signal-to-noise ratios. Again, the L-MBBP approach mitigates the gap to the random coding bound by about $20$ $\%$. 

It is worth mentioning that the codes defined in the WiMAX standard have a significantly lower density compared to the considered PEG codes. This allows for faster decoding with the BP algorithm. If one considers not only the length but also the decoding speed as a system parameter, the standardized codes are comparable to the PEG-optimized codes discussed in this work. Detailed results on this comparison can be found in \cite{hehn09}.

\section{Conclusions}

The contribution of this paper is two-fold. First, we adapted the scheme of MBBP decoding to codes from the WiMAX standard what allowed us to improve the decoding performance for about $0.15$ dB. As a second contribution, we compared the performance of the WiMAX codes with the performance of PEG codes. Both for BP decoding and MBBP decoding, the PEG codes obtained a measurable performance improvement compared to the codes in the WiMAX standard. 

\bibliography{LDPC_Group_Bibfile}
\bibliographystyle{IEEEtran}

\end{document}